\documentclass[10pt,twocolumn,letterpaper]{article}

\usepackage[pagenumbers]{cvpr}      
\definecolor{cvprblue}{rgb}{0.21,0.49,0.74}
\usepackage[pagebackref,breaklinks,colorlinks,allcolors=cvprblue]{hyperref}
\usepackage{hyperref}
\usepackage{graphicx}
\graphicspath{{./images/}}
\usepackage{multirow}

\usepackage{algorithm}
\usepackage{alphalph}
\usepackage{algorithmic}
\usepackage{bbding}
\usepackage{makecell}
\usepackage[accsupp]{axessibility}

\usepackage{orcidlink}

\usepackage{enumitem}

\usepackage{subcaption}

\usepackage[utf8]{inputenc} 
\usepackage[T1]{fontenc}    
\usepackage{url}            
\usepackage{booktabs}       
\usepackage{amsfonts}       
\usepackage{nicefrac}       
\usepackage{microtype}      
\usepackage{amssymb}
\usepackage{pifont}

\usepackage[table,dvipsnames]{xcolor}
\definecolor{LightCyan}{rgb}{0.91,0.91,0.98}
\definecolor{LightYellow}{rgb}{1.0, 1.0, 0.88}
\definecolor{magicmint}{rgb}{0.67, 0.94, 0.82}
\definecolor{lightmauve}{rgb}{0.86, 0.82, 1.0}
\definecolor{grannysmithapple}{rgb}{0.66, 0.89, 0.63}
\definecolor{isabelline}{rgb}{0.95,0.93,0.91}

\def\paperID{16367} 
\def\confName{CVPR}
\def\confYear{2026}

\title{APRVOS: 1st Place Winner of 5th PVUW MeViS-Audio Track}
\author{
Deshui Miao$^{1}$ \quad
Yameng Gu$^{1}$ \quad
Chao Yang$^{1}$ \\
Xin Li$^{1*}$ \quad
Haijun Zhang$^{2}$ \quad
Ming\mbox{-}Hsuan Yang$^{3}$ \\
$^{1}$Pengcheng Laboratory \quad
$^{2}$Harbin Institute of Technology \\
$^{3}$University of California at Merced 
}

\begin{document}
\maketitle
\begin{abstract}
This report presents an Audio-aware Referring Video Object Segmentation (Ref-VOS) pipeline tailored to the MEVIS\_Audio setting, where the referring expression is provided in spoken form rather than as clean text. Compared with a standard Sa2VA-based Ref-VOS pipeline, the proposed system introduces two additional front-end stages: speech transcription and visual existence verification. Specifically, we first employ VibeVoice-ASR to convert long-form spoken input into a structured textual transcript. Since audio-derived queries are inherently noisy and may describe entities that are not visually present in the video, we then introduce an Omni-based judgment module to determine whether the transcribed target can be grounded in the visual content. If the target is judged to be absent, the pipeline terminates early and outputs all-zero masks. Otherwise, the transcript is transformed into a segmentation-oriented prompt and fed into Sa2VA to obtain a coarse mask trajectory over the full video. Importantly, this trajectory is treated as an initial semantic hypothesis rather than a final prediction. On top of it, an agentic refinement layer evaluates query reliability, temporal relevance, anchor quality, and potential error sources, and may invoke SAM3 to improve spatial boundary precision and temporal consistency. The resulting framework explicitly decomposes the MEVIS\_Audio task into audio-to-text conversion, visual existence verification, coarse video segmentation, and agent-guided refinement. Such a staged design is substantially more appropriate for audio-conditioned Ref-VOS than directly sending noisy ASR outputs into a segmentation model.
\end{abstract}

\section{Introduction}
Referring Video Object Segmentation (Ref-VOS)~\cite{ding2023mevis,liu2023gres,khoreva2019video,pont20172017} aims to segment, throughout a video, the object or entity specified by a referring expression. In conventional Ref-VOS benchmarks~\cite{cheng2022xmem,wu2022language,gong2025reinforcing,videolisa,zheng2024villa}, the query is usually provided as a clean and unambiguous text string. MEVIS\_Audio is fundamentally different: the referring signal is delivered as speech. As a result, the system must not only solve visual grounding and mask prediction, but also first recover the actual linguistic intent from an audio stream.
CVPR 2026 5th PVUW challenge has three tracks: Complex VOS on
MOSEv2~\cite{ding2025mosev2}, targeting realistic, cluttered scenes with
small, occluded, reappearing, and camouflaged objects
under adverse conditions; VOS on MOSE~\cite{ding2023mose}, focusing
on challenging, long, and diverse videos; and RVOS on
MeViS~\cite{ding2023mevis, MeViSv2}, assessing referring video object segmentation with text or audio.

This difference introduces several additional sources of uncertainty. First, automatic speech recognition (ASR) errors can directly distort the semantic content of the query, leading to downstream failures in grounding~\cite{zhang2024ferret,wang2025time,bai2025univg,yang2023set} and segmentation~\cite{hyperseg,luo2024soc,liu2024universal}. Second, spoken referring expressions are typically less well-formed than written ones: they often contain hesitations, filler words, repetitions, incomplete noun phrases, delayed references, and colloquial descriptions. Third, even when the transcribed content is linguistically meaningful, the mentioned target may still be absent from the visual scene. Therefore, directly reducing MEVIS\_Audio to a standard text-conditioned Ref-VOS problem is not sufficiently robust.
\begin{table*}[t]
\centering
\caption{Leaderboard results of MeViS\_Audio track.}
\label{tab:leaderboard}
\begin{tabular}{c l c c c c c c}
\toprule
Rank & Participant & J\&F & J & F & N-acc. & T-acc. & Final \\
\midrule
1 & Ours       & 0.6700 & 0.6381 & 0.7019 & 0.8939 & 0.9767 & 0.846857 \\
2 & wangzhiyu918 & 0.6387 & 0.6098 & 0.6675 & 0.8333 & 0.9494 & 0.807134 \\
3 & csjihwanh    & 0.5394 & 0.5159 & 0.5630 & 0.6970 & 0.8157 & 0.684025 \\
4 & vvv666       & 0.4716 & 0.4406 & 0.5025 & 0.1212 & 0.9767 & 0.523139 \\
5 & liyiying     & 0.4769 & 0.4490 & 0.5048 & 0.0909 & 0.9650 & 0.510930 \\
\bottomrule
\end{tabular}
\end{table*}

The central idea of our MEVIS\_Audio pipeline is to explicitly decouple three decisions that are often conflated in a single-pass model: \emph{(i)} what the speaker actually said, \emph{(ii)} whether the spoken target exists in the video, and \emph{(iii)} how the target should be segmented and refined over time. To this end, we design a staged pipeline. VibeVoice-ASR~\cite{peng2026vibevoice} first transcribes the spoken input into text. An Omni-based judgment module then determines whether the transcribed target is visually present. Only if this existence check succeeds do we invoke Sa2VA~\cite{yuan2025sa2va} to produce a coarse mask trajectory. Finally, an agentic refinement stage evaluates the reliability of the coarse prediction, identifies trustworthy anchor frames, and optionally calls SAM3~\cite{carion2025sam} to obtain sharper and more temporally stable masks.

This report focuses specifically on the MEVIS\_Audio variant of the method. Relative to the text-only Ref-VOS~\cite{yan2024visa,lin2025glus,li2024omg,kirillov2023segment} pipeline, its defining extension is the addition of a \texttt{VibeVoice-ASR + Omni judgment} front-end, which is essential for handling audio-originated uncertainty before dense segmentation begins.

\section{Problem Setup}
Let the input consist of a video
\begin{equation}
V = \{I_t\}_{t=1}^{T}
\end{equation}
and an audio referring signal
\begin{equation}
A.
\end{equation}
The goal is to predict a binary mask sequence
\begin{equation}
\mathcal{M} = \{m_t\}_{t=1}^{T}, \qquad m_t \in \{0,1\}^{H_t \times W_t},
\end{equation}
where each $m_t$ denotes the segmentation mask of the target object at frame $t$.

Since the referring query is not directly available in textual form, we first perform speech transcription:
\begin{equation}
q_{\text{asr}} = \mathrm{VibeVoice}(A),
\end{equation}
where $q_{\text{asr}}$ denotes the transcript or cleaned expression candidate extracted from the spoken input. Next, we estimate whether the transcribed target is visually present in the video:
\begin{equation}
e = \mathrm{Omni}(V, q_{\text{asr}}), \qquad e \in \{0,1\},
\end{equation}
where $e=1$ indicates that the target is judged to exist in the video and $e=0$ otherwise.

The final prediction rule is defined as
\begin{equation}
\mathcal{M} =
\begin{cases}
\{\mathbf{0}\}_{t=1}^{T}, & e = 0, \\
\Phi(V, q_{\text{asr}}), & e = 1,
\end{cases}
\end{equation}
where $\Phi$ denotes the downstream segmentation-and-refinement procedure built upon Sa2VA and the agentic post-processing module.

This formulation is directly consistent with the current Sa2VA~\cite{yuan2025sa2va} evaluation flow for MEVIS\_Audio. In particular, the dataset split is associated with \texttt{meta\_expressions\_audio\_asr.json}, meaning that the query is explicitly ASR-derived, while \texttt{presence\_info.target\_exists} provides a natural interface for the early-stop decision. Hence, the MEVIS\_Audio setting is not merely a change in data format; it explicitly requires modeling both speech-to-text uncertainty and target-existence uncertainty before segmentation.

\section{Method}
\begin{figure*}
    \centering
    \resizebox{\linewidth}{!}{
        \includegraphics{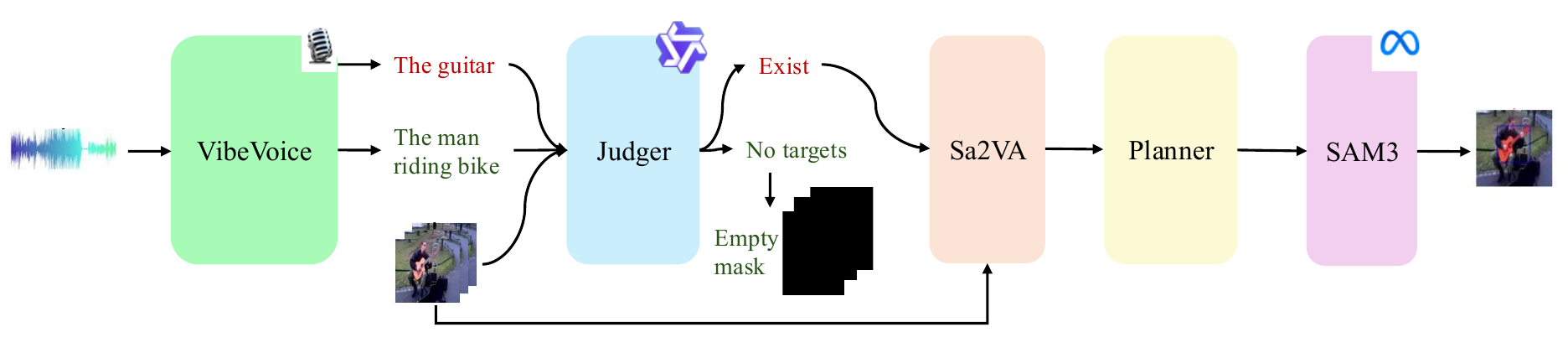}
    }
    \caption{\textbf{Pipeline of our methods.}}
    \label{fig:mevis_audio}
     \vspace{-4mm}
\end{figure*}

\subsection{Stage -1: VibeVoice-ASR~\cite{peng2026vibevoice} for Audio-to-Text Conversion}
The first additional component in the MEVIS\_Audio pipeline is VibeVoice-ASR~\cite{peng2026vibevoice}, whose role is to convert the spoken referring signal into a machine-readable textual representation. This stage is particularly important because the referring phrase may appear inside a longer utterance rather than as a short, isolated command. VibeVoice-ASR~\cite{peng2026vibevoice} is suitable for this setting because it is designed for long-form speech recognition and can provide structured transcriptions that preserve useful contextual cues, such as speaker turns, temporal alignment, and utterance content.

Crucially, this stage does not perform any visual reasoning or segmentation. Its sole purpose is to recover the semantic content of the spoken query as faithfully as possible. In practice, it may also support ambiguity reduction through lexical priors, hotwords, or domain-specific cues, which is beneficial when the speaker refers to uncommon object names, personal nicknames, or task-specific terminology. The output of this stage is a transcript or cleaned expression candidate that can be stored in \texttt{meta\_expressions\_audio\_asr.json} and subsequently consumed by the Ref-VOS pipeline.

Conceptually, this stage transforms an uncertain acoustic signal into a textual hypothesis. Because every downstream step relies on this hypothesis, explicit ASR modeling is indispensable in MEVIS\_Audio and should not be treated as a negligible preprocessing detail.

\subsection{Stage 2: Visual Existence Judgment}
After obtaining the ASR transcript, the next stage performs visual existence judgment. Its purpose is not to predict dense masks, but to answer a binary and highly consequential question: \emph{does the transcribed target actually appear in the video?}

To address this question, we employ Qwen3-VL~\cite{Qwen2.5-VL} as a visual judge. Given the transcript-derived referring phrase and a set of sampled video frames, the module estimates whether the described entity can be visually grounded in the scene. The result is stored as \texttt{presence\_info.target\_exists}.

This stage serves as an essential robustness mechanism against ASR-induced false positives. A transcribed phrase may be linguistically plausible while still having no corresponding visual instance in the video. Without an explicit existence gate, the downstream segmenter would be forced to hallucinate a mask for a nonexistent target, thereby degrading both semantic validity and evaluation performance. By inserting this binary verification stage before dense segmentation, the pipeline avoids unnecessary computation and reduces error propagation from the audio front-end.

This design also aligns naturally with the current Sa2VA~\cite{yuan2025sa2va} Ref-VOS evaluation script. When \texttt{target\_exists} is false, the runner emits the text prediction \texttt{[META:NO\_OBJ] target\_exists=false} and outputs all-zero masks for the entire sequence. Therefore, the front-end of the MEVIS\_Audio pipeline can be cleanly summarized as
\begin{equation}
\texttt{audio} \rightarrow \texttt{ASR transcript} \rightarrow \texttt{
 Existence}.
\end{equation}

\subsection{Stage 3: Prompt Construction for Sa2VA~\cite{yuan2025sa2va}}
If the target is judged to be present, the transcript is converted into a segmentation-oriented textual prompt that matches the input format expected by Sa2VA~\cite{yuan2025sa2va}. In the current dataset loader, this is achieved through one of two prompt templates:
\begin{enumerate}[leftmargin=1.3em]
    \item \texttt{<image>\textbackslash nPlease segment \{exp\}.}
    \item \texttt{<image>\textbackslash n\{exp\} Please respond with a segmentation mask.}
\end{enumerate}
The second template is used when the expression already resembles a question.

Although seemingly simple, this prompt-construction step is functionally important. The ASR output is not automatically suitable as a segmentation query. It must be transformed into a textual instruction whose format is compatible with the multimodal interface of the downstream Ref-VOS model. In other words, this stage bridges the gap between raw speech transcription and segmentation-conditioned language prompting.

\subsection{Stage 4: Coarse Semantic Segmenter}
The constructed prompt, together with the full video, is then fed into Sa2VA. Let
\begin{equation}
\tilde{\mathcal{M}} = \{\tilde{m}_t\}_{t=1}^{T} = \mathrm{Sa2VA}(V, q_{\text{asr}})
\end{equation}
denote the coarse mask trajectory returned by \texttt{predict\_forward}. Since Sa2VA processes the entire video and outputs \texttt{prediction\_masks} aligned with the original frame sequence, it functions as a full-video segmenter rather than a frame-local grounding model.

This stage provides the first dense, semantically grounded estimate of the target trajectory. However, in the MEVIS\_Audio setting, we do not treat this output as immediately final. There are two main reasons. First, residual ASR noise may still contaminate the semantics of the prompt, leading to imperfect grounding. Second, even when the transcript is correct, Sa2VA may produce masks with coarse object boundaries, occasional distractor confusion, or temporally unstable behavior. Therefore, the Sa2VA prediction should be interpreted as a \emph{semantic prior} or \emph{initial hypothesis} rather than a definitive answer.

This distinction is important to the overall system design: Sa2VA provides broad semantic coverage over the video, while later stages determine which parts of its output are reliable enough to preserve and which parts require correction or refinement.

\subsection{Stage 3: Agentic Verification}
To improve robustness beyond single-shot segmentation, we introduce an agentic reasoning layer on top of the coarse Sa2VA output. Rather than blindly accepting the predicted mask trajectory, this layer explicitly evaluates its reliability from multiple perspectives. For example, it can inspect which frames contain non-empty masks, whether the mask area changes smoothly over time, whether the predicted object remains semantically consistent with the spoken description, and whether multiple visually similar distractors may have caused grounding ambiguity.

This stage is where the complexity of audio-conditioned Ref-VOS is handled through explicit reasoning rather than being hidden inside one end-to-end segmentation score. The agent can analyze the transcript quality, infer the most relevant temporal window, identify candidate anchor frames, and decompose the query into positive constraints, negative constraints, and temporal hints. A planner may decide which refinement strategy is most appropriate; scout modules may search for frames in which Sa2VA provides the most trustworthy localization signal; and a critic may assess whether the current trajectory is semantically plausible and temporally coherent.

Such an agentic layer is especially valuable in MEVIS\_Audio because uncertainty enters from both the language side and the visual side. By explicitly reasoning about these uncertainties, the system gains a mechanism for selective trust, targeted correction, and failure-aware refinement.

\subsection{Stage 4: Refinement from Trusted Anchors}
Once the agent identifies a reliable anchor frame, the corresponding Sa2VA mask can be converted into geometric prompts for SAM3-based refinement. For a trusted frame $a$, we derive
\begin{equation}
b_a = \mathrm{BBox}(\tilde{m}_a), \qquad
p_a = \mathrm{Center}(\tilde{m}_a),
\end{equation}
or, if needed, an alternative refinement point predicted by a visual-language model. These geometric prompts are then used to initialize SAM3, which propagates the target both forward and backward in time.

This refinement stage improves the prediction in two complementary aspects. Spatially, SAM3 can recover sharper object boundaries than the coarse masks produced by the initial segmenter. Temporally, anchor-based propagation can stabilize the object trajectory and reduce frame-to-frame inconsistency. As a result, Sa2VA and SAM3 play distinct but complementary roles: Sa2VA contributes global semantic grounding, while SAM3 provides high-quality boundary recovery and propagation once a trustworthy initialization has been identified.


\section{Experiments}

We report a simple ablation study on MEVIS\_Audio to measure the incremental value of each stage. Since the purpose of this report is to explain the method design rather than a full benchmark protocol, we present a single overall score for each variant. The comparison starts from plain Sa2VA and then adds the proposed components one by one.

\begin{table}[h]
\centering
\begin{tabular}{p{0.72\linewidth}c}
\hline
\textbf{Method} & \textbf{Score} \\
\hline
Sa2VA-4B without judgment & 0.45 \\
Sa2VA-26B without judgment & 0.53 \\
Sa2VA-4B + Omni judgment & 0.55 \\
Sa2VA-4B + Omni judgment + SAM3 refine & 0.59 \\
Sa2VA-4B + Omni judgment + SAM3 refine + planner + SA~\cite{yuan2025sa2va} & \textbf{0.67} \\
\hline
\end{tabular}
\caption{All results on MEVIS\_Audio.}
\end{table}

Three observations are immediate. First, simply scaling Sa2VA from 4B to 26B improves the score from 0.45 to 0.53, which confirms that model capacity helps, but only to a limited extent. Second, adding the judgment stage to the 4B model already raises the score to 0.55, exceeding the raw 26B baseline. This indicates that filtering out visually absent or semantically mismatched audio expressions is more important than only increasing the backbone size. Third, refinement and planning provide additional gains: SAM3-based refinement raises the score to 0.59, and adding the planner pushes the result to 0.67, which is the best result in this sequence.

These numbers support the central claim of the report. MEVIS\_Audio is not merely a bigger-model problem. Its difficulty comes from error accumulation across speech recognition, existence judgment, dense grounding, and temporal refinement. Once the pipeline is organized explicitly as \texttt{VibeVoice-ASR} $\rightarrow$ \texttt{Omni judgment} $\rightarrow$ \texttt{Sa2VA} $\rightarrow$ \texttt{SAM3 refine} $\rightarrow$ \texttt{planner}, the score improves much more consistently than by scaling Sa2VA alone.

\section{Conclusion}
This report introduced a Ref-VOS pipeline specifically designed for the MEVIS\_Audio setting by augmenting a Sa2VA-based framework with two additional front-end stages: VibeVoice-ASR for speech transcription and Omni-based judgment for target-existence verification. The overall procedure can be summarized as
\begin{equation}
\begin{aligned}
\text{audio}
&\rightarrow \text{VibeVoice-ASR}
\rightarrow \text{Judgment} \\
&\rightarrow \text{Coarse masks}
\rightarrow \text{Agentic refinement} \\
&\rightarrow \text{SAM3 sharpening}.
\end{aligned}
\end{equation}

The key insight is that MEVIS\_Audio should not be treated as an ordinary text-conditioned segmentation problem. A correct solution requires explicit modeling of both speech-recognition noise and visual-existence uncertainty before dense segmentation begins. By decomposing the task into transcription, verification, coarse grounding, and refinement, the proposed pipeline offers a more principled and robust formulation for audio-conditioned Ref-VOS.

{
    \small
    \bibliographystyle{ieeenat_fullname}
    \bibliography{main}
}


\end{document}